# Mental Health Pandemic during the COVID-19 Outbreak:

# Social Media As a Window to Public Mental Health

Michelle Bak[1], Chungyi Chiu[2], and Jessie Chin[1]

[1] School of Information Sciences, University of Illinois Urbana-Champaign

[2] Department of Kinesiology and Community Health, University of Illinois Urbana-Champaign

## Author Note

Correspondence concerning this article should be addressed to Michelle Bak, School of Information Sciences, University of Illinois at Urbana-Champaign, 614 East Daniel Street, Champaign, IL, 61820, USA. Email: chaewon7@illinois.edu

Abstract

Intensified preventive measures during the COVID-19 pandemic, such as lockdown and social distancing, heavily increased the perception of social isolation (i.e., a discrepancy between one's social needs and the provisions of the social environment) among young adults. Social isolation is closely associated with situational loneliness (i.e., loneliness emerging from environmental change), a risk factor for depressive symptoms. Prior research suggested vulnerable young adults are likely to seek support from an online social platform such as Reddit, a perceived comfortable environment for lonely individuals to seek mental health help through anonymous communication with a broad social network. Therefore, this study aims to identify and analyze depression-related dialogues on loneliness subreddits during the COVID-19 outbreak, with the implications on depression-related infoveillance during the pandemic. Our study utilized logistic regression and topic modeling to classify and examine depression-related discussions on loneliness subreddits before and during the pandemic. Our results showed significant increases in the volume of depression-related discussions (i.e., topics related to mental health, social interaction, family, and emotion) where challenges were reported during the pandemic. We also found a switch in dominant topics emerging from depression-related discussions on loneliness subreddits, from dating (prepandemic) to online interaction and community (pandemic), suggesting the increased expressions or need of online social support during the pandemic. The current findings suggest the potential of social media to serve as a window for monitoring public mental health. Our future study will clinically validate the current approach, which has implications for designing a surveillance system during the crisis.

## Introduction

The current Coronavirus Disease 2019 (COVID-19) has heavily affected normalcy globally.[1] On March 11, 2020, the World Health Organization (WHO) declared the COVID-19 outbreak a pandemic.[2] Although following the protective measures is critical to restrict the spread of the virus, lockdown and social distancing have escalated the perceived social isolation (i.e., a discrepancy between one's social needs and the actual or perceived provisions of the social environment),[3] particularly among young adults.[4] Young adults are at their emerging adulthood, a developmental stage between adolescence and adulthood from approximately the late teens through the 20s.[5]

One of the prominent features of emerging adulthood is the development or intensification of identity exploration (i.e., clarification process of one's sense of self), which occurs through interacting with different individuals in diverse social environments.[6] Although abundant opportunities for social relationships are closely related to the positive development of young adults,[7] the perceived social isolation induced by the pandemic puts young adults at higher risks of situational loneliness (i.e., a type of loneliness emerging from environmental change),[4,8] a risk factor for depressive symptoms.[9]

Indeed, the percentage difference in depression levels between the prepandemic and pandemic periods was the greatest among younger adults aged 18–34 years compared with other sociodemographic groups.[10] Given that depression, a mental disorder characterized by mild-to-severe emotional and physical symptoms (e.g., persistent sadness, feelings of hopelessness, or increased fatigue) lasting for at least 2 weeks,[11] is associated with substance use disorders,[12] higher risks of other chronic health conditions,[13] or suicidal ideation and attempt,[14] the prevalence of depression among young adults during the pandemic[15] is particularly alarming.

Understanding their depression-related views or experiences emerging from loneliness dialogues is critical to address such a mental health crisis. Hence, this study aims to investigate depression-related discussions on Reddit, a discussion-based online community[16] primarily composed of young adult users.[17] Reddit consists of different subgroups called “subreddits” dedicated to a single topic, and users can choose to join subreddits of their interests.[18]

Reddit users interact with others by submitting a post, commenting on a post,[16] and upvoting or downvoting to express their support (and endorsement) or disagreement.[19] Posts that receive a higher number of upvotes become more visible.[20] Reddit is the fifth most frequently visited Web site in the United States,[21] with users located in the United States as the dominant Reddit users.

We chose Reddit for two primary reasons—user demographics and anonymity practices on Reddit. First, the highest Reddit use was reported in the users aged 18 to 29 years (36 percent) among other age groups (3–22 percent).[17] In addition, Reddit does not require user self-identification for registration, providing the users an option to practice anonymity on the platform (i.e., throwaway accounts).[22,23] Prior research suggested individuals experienced with loneliness are likely to have low self-esteem[24] when perceiving a high level of public self-consciousness.[25–27] In addition, they are likely to seek help from an online social platform, a perceived comfortable environment for lonely people to seek mental health support through anonymous communication with a broad social network.[28–30]

For example, a previous study showed that individuals who expressed loneliness are likely to seek support and advice from online loneliness forums such as Reddit.[31] Indeed, Reddit is widely utilized by individuals with mental health concerns as it provides an anonymous environment for users who are hesitant to engage in self-disclosure due to the stigma about mental illness.[32] Hence, this study strives to identify and analyze depression-related dialogues emerging during the COVID-19 outbreak from loneliness subreddits (r/lonely, r/loneliness, and r/ForeverAlone),[33–35] primarily composed of young adults.

Young adults are likely to experience high levels of situational loneliness induced by social isolation during the pandemic.[3,4,8] Given the high risk of depressive symptoms in individuals experiencing loneliness,[9] we first built a classifier that detects depression-related discussions from loneliness subreddits. Then, we examined the content of depression-related posts (DPs) on loneliness subreddits to understand their experiences during the pandemic.

## Study 1: Building a DP Classifier

We utilized inductive transfer learning to build a classifier that detects DPs from loneliness subreddits.[36] Inductive transfer learning utilizes the knowledge obtained from learning a different but related task to improve performance on the target task.[37] This approach is commonly adopted to detect mental health expressions on social media.[38]

### Data collection

To build a binary classifier that detects DPs, we collected the title and body text of posts from the largest depression subreddit, r/depression[39] (N = 99,999; target task) and nondepression subreddit, r/NoStupidQuestions[40] (N = 100,000; different but related task) using pushshift.io Reddit API.[41] According to the US Department of Health and Human Services, this study is IRB exempt due to the use of secondary data analysis on publicly available data.[42]

### Data preprocessing

We dropped duplicate rows or rows where body text was missing, deleted, or removed. This reduced the data size of r/depression and r/NoStupidQuestions to 52,852 and 42,614, respectively. We then aggregated the title and body text into a single text for each entry. Because a balanced training data set (i.e., equal distribution of labels presented in the training data set) is important to build an unbiased classifier, we randomly sampled 42,614 rows from r/depression to be equivalent to the size of data obtained from r/NoStupidQuestions. We then labeled data from r/depression as depression-related and r/NoStupidQuestions as unrelated to depression. These data sets were merged and used for training a classifier that detects DPs.

**Building a classifier**

The text data set was split into training (80 percent) and testing set (20 percent) and transformed into a matrix of token counts using CountVectorizer.[43] We trained the text data using supervised classification algorithms extensively employed for detecting DP, such as logistic regression,[44,45] support vector machine,[46] multinomial naive Bayes,[47,48] and random forest.[49,50] A 10-fold cross-validation method was leveraged to evaluate the model performances.[51]

**Results**

Table 1 gives the performance of each model in predicting DPs. We evaluated the models using four common performance metrics in machine learning,[52] the accuracy, precision, recall, and F1 scores (see definitions in Table 1).[52,53] We found logistic regression achieved the highest accuracy, 0.95 and F1 score, 0.95. Supplementary Figure S1 demonstrates a confusion matrix of the best performing model. In addition, the area under the curve value of receiver operating characteristic[54–56] yielded 0.98 (Supplementary Fig. S2). All the metrics demonstrated the excellent performance of the model.

**Table 1**. Prediction Results for classifying depression-related posts

| Classifier | F1 score | accuracy | precision | recall |
|---|---|---|---|---|
| Logistic Regression | 0.95 | 0.95 | 0.95 | 0.94 |
| Support Vector Machine | 0.94 | 0.94 | 0.95 | 0.94 |
| Multinomial Naïve Bayes | 0.93 | 0.93 | 0.89 | 0.97 |
| Random Forest | 0.91 | 0.91 | 0.90 | 0.93 |

Note. *accuracy* is computed by dividing the sum of true positives (TP) and true negatives (TN) by the total number of predictions, *precision* is computed by dividing TP by the sum of TP and false positives (FP), *recall* is computed by dividing TP by the sum of TP and false negatives (FN), and *F1 scores* is the harmonic mean between precision and recall.[52,53]

**Supplementary Figure 1**. Confusion matrix for logistic regression. "dep" means related to depression and "non-dep" mean not related to depression. TN=true negative; FP=false positive; FN=false negative; TP=true positive

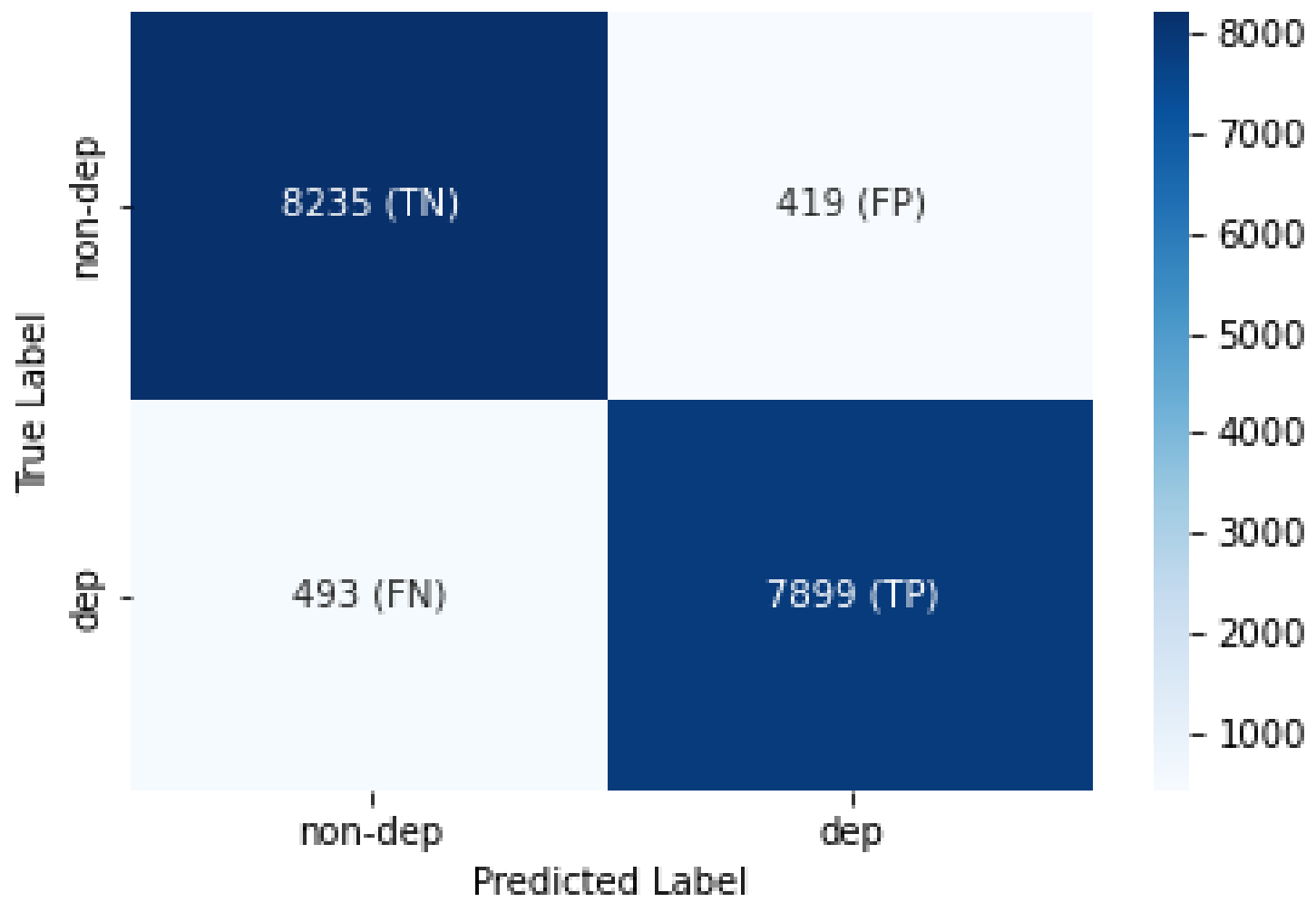

**Supplementary Figure 2**. ROC for logistic regression

*Note*. AUC value of receiving operating characteristic (ROC) is true positive rates (TP / (TP+FN))[65-66] against the false positive rates (FP / (FP+TN))[67]

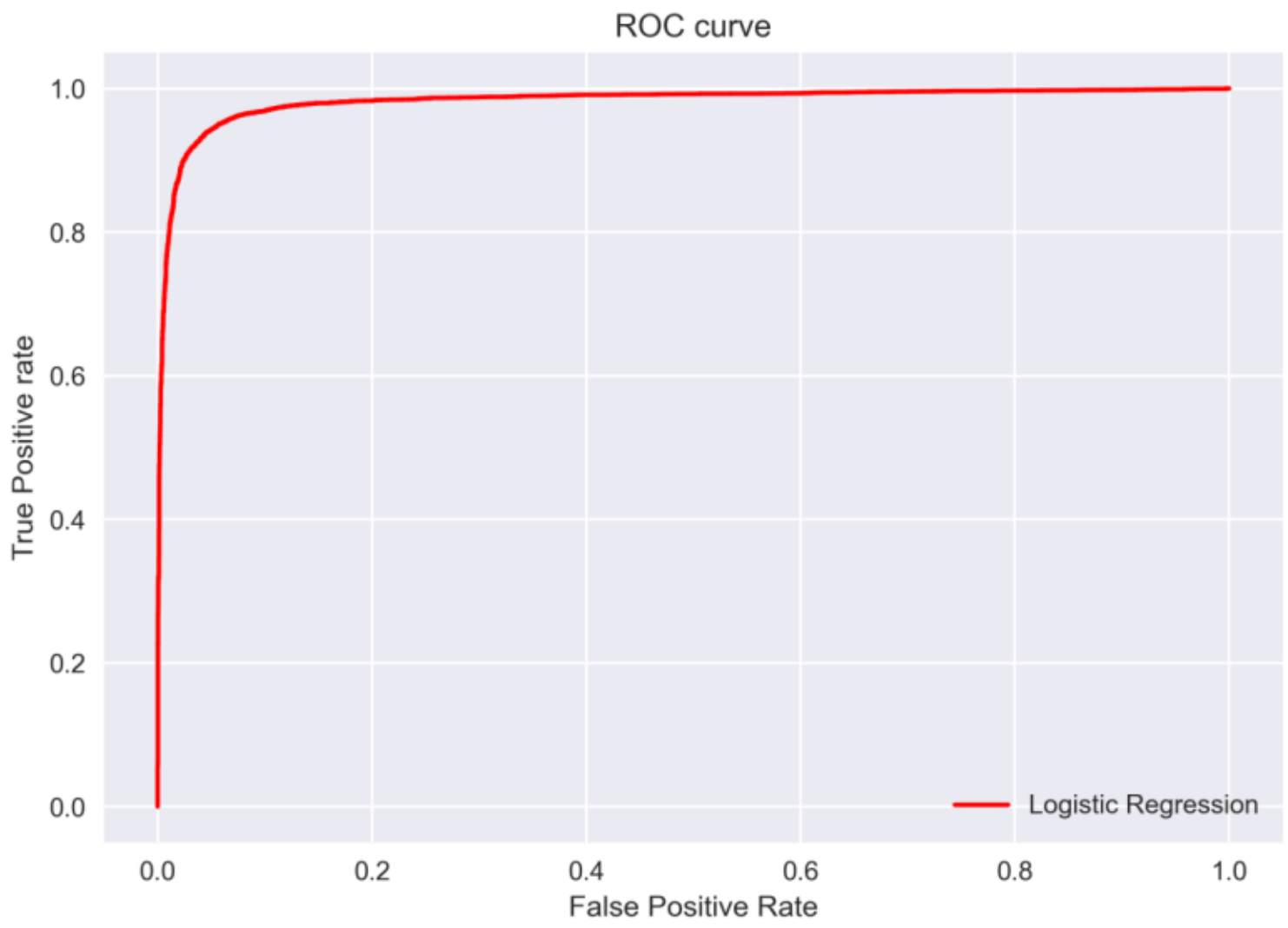

## Study 2: Detecting DPs from Loneliness Subreddits

### Data Collection

We collected post content and associated metadata from the largest loneliness-related subreddits, r/lonely, r/loneliness, and r/ForeverAlone. To compare DPs on loneliness subreddits posted before and during the pandemic, we collected Reddit data using two time periods, from March 11, 2018, to September 22, 2019 (T1) and March 11, 2020, and September 22, 2021 (T2), respectively. Table 2 gives the number of data collected from each subreddit across two time periods.

**Table 2**. Number of data collected from each loneliness-related subreddit

| Name of subreddit | T1 | T2 | Percentage increase |
|---|---|---|---|
| r/lonely | N=19,767 | N=26,530 | 34.21% |
| r/loneliness | N=1,054 | N=1,985 | 88.33% |
| r/ForeverAlone | N=27,988 | N=67,450 | 141.00% |
| | Total N=48,809 | Total N=95,965 | 96.61% |

### Data preprocessing

Data preprocessing of loneliness subreddits data was identical to that of Study 1. This reduced the total data size from 48,809 to 17,885 (T1), and from 95,965 to 41,398 (T2).

### Applying a classifier

A classifier from Study 1 was applied to the preprocessed data.

### Results

Among the posts from the loneliness subreddits during T1 (N = 17,885), 69.16 percent of posts (N = 12,372) included DP (Table 3). Among T2 posts (N = 41,398), 80.49 percent of posts (N = 33,321) contained DP. A chi-square test revealed a significant difference between the two proportions (chi-square = 904.82; $p < 0.05$). Similarly, Table 4 demonstrates a significant increase in the number of unique users who mentioned DPs on loneliness subreddits from T1 to T2 (chi-square = 95.42; $p < 0.05$). These results suggested an increasing DP during the pandemic on social media, which resonates with an increased number of depression cases during the pandemic.[57–59]

**Table 3**. Number of posts

| | T1 | T2 | chi-square |
|---|---|---|---|
| Number of posts on loneliness subreddits | N=17,885 | N=41,398 | 904.82* |
| Number of posts that mention depression-related topics on loneliness subreddits | N=12,372 | N=33,321 | |

*p<0.05

**Table 4**. Number of unique users

| | T1 | T2 | chi-square |
|---|---|---|---|
| Number of unique users on loneliness subreddits | N=7,557 | N=22,932 | 95.42* |
| Number of unique users who discuss depression-related topics on loneliness subreddits | N=6,169 | N=19,778 | |

*p<0.05

## Study 3: Analyzing Depression-Related Posts (DP) on Loneliness Subreddits

We used the topic modeling algorithm latent Dirichlet allocation (LDA)[60] to examine and compare topics of DPs in loneliness subreddits between two time periods. LDA is a generative probabilistic model that automatically discovers hidden topics from text documents. LDA considers a document as a random mixture of latent topics where each topic is characterized by a distribution of word probabilities.

### Data Preprocessing

We first merged T1 and T2 posts and removed white spaces and quotes from the combined posts. Then, we excluded stopwords (e.g., the, a) and lemmatized the text data to filter terms by parts of speech using Python library SpaCy.[61] We calculated term frequency-inverse document frequency representations of text data[62] using Gensim, a Python library.[63] This final corpus was used to fit our topic model.

### Topic Model Evaluation

We fitted the LDA models with the Machine Learning for Language Toolkit (MALLET) wrapper of Gensim[64] and calculated a set of coherence scores using the number of topics (k) from 5 to 50.[65] We then identified the optimal number of topics (k = 17) guided by the topic coherence value (Supplementary Fig. S3). Using pyLDAvis, we visualized our topic model to interpret the topic distributions and their associated terms ranked by probabilities.[66–68] Two annotators labeled the topics independently, and annotation discrepancies were resolved through discussions.

**Supplementary Figure 3**. Coherence scores for topic models (highest value at k=17)

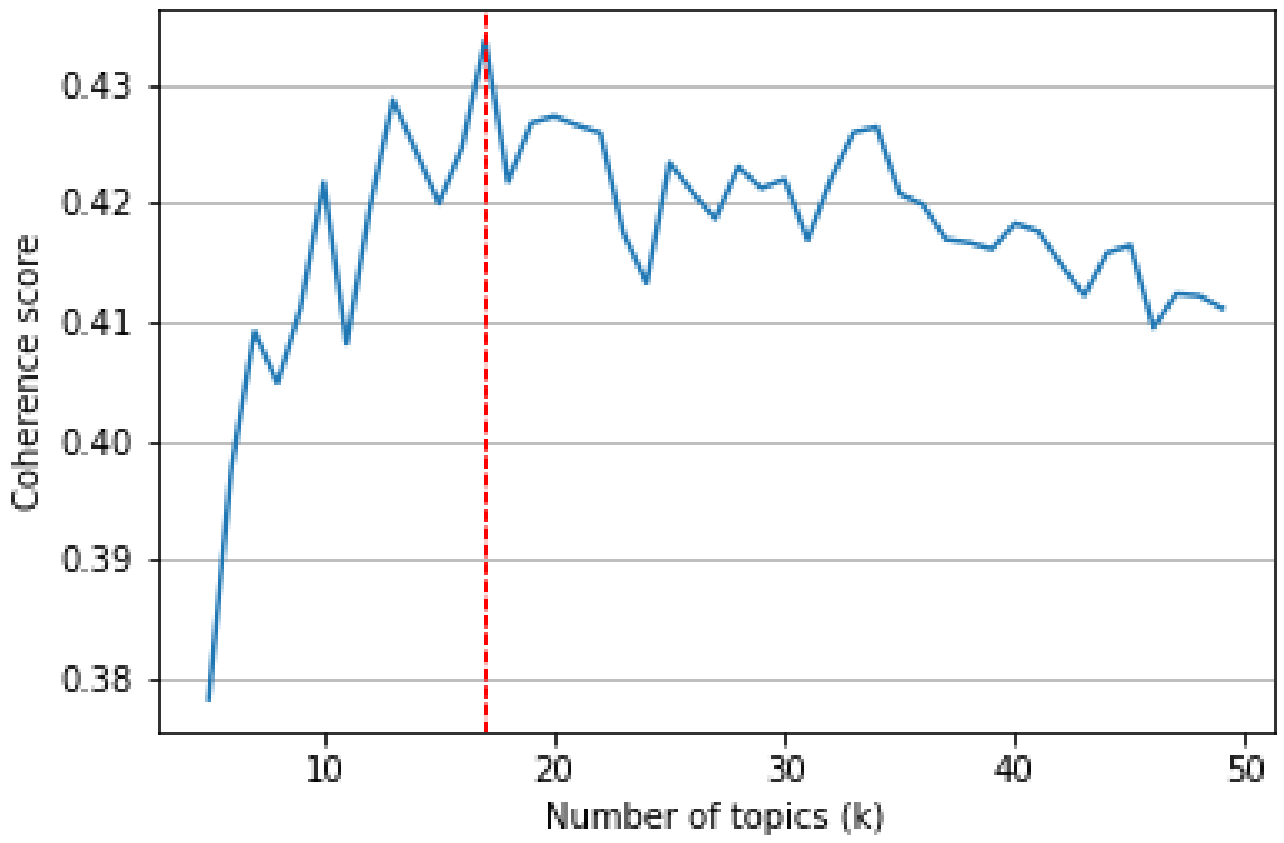

## Results

A topic model of posts revealed 17 topics categorized into 9 broad themes: mental health, social interaction, family, emotion, job, hobby, college life, human mind, and time period (Tables 5 and 6). Under the mental health category, significant increases in DP on experiencing and venting negative emotions align with the elevated mental health risks during the pandemic, given the increased perceived social isolation and loneliness.[3,69] No change in DPs on mental health conditions was found in loneliness subreddits.

**Table 5**. Topics and associated keywords

| Topic Label | Sample Keywords | Sample Post |
|---|---|---|
| Mental health conditions | depression, mental, anxiety, therapy, mentally, struggle, suicidal | It feels like nothing is real from time to time. Is this because of my recurring depression, anxiety, or am I just slowly going insane? |
| Experiencing negative emotions | lonely, happy, sad, loneliness, feeling, hard, depressed, bad, empty, hurt, scared, sadness | I feel sad all the time and I cry myself to sleep. I hate myself and everyone, and disdain happy people around me. |
| Venting negative emotions | tired, hate, stupid, sick, pathetic, end, worthless, dead, miserable, loser | I don't know if I can continue like this. I knew s**t was going to get worse, but it shouldn't be. |
| Relationship | relationship, close, single, person, couple, boyfriend, partner, romantic, friendship | I don't have any friends and recently broke up with my boyfriend few months ago. This makes me depressed and numb, questioning relationships. |
| Social evaluation | people, person, care, hard, kind, nice, attention, effort, fake, selfish, honest, trust | Why is society so brutal and uncaring? People are nasty and don't care about each other, taking advantage of each other. |
| Social experience | social, connection, normal, experience, awkward, interaction, anxiety, weird | It is difficult for me to maintain friendships and I feel lonely. I never had a chance to develop proper social skills, making me feel like a total outcast. |
| Dating | girl, guy, woman, date, man, sex, girlfriend, pretty, single, app, attractive, rejection | How can I be happy if no one wants to ever be with me? I feel like no matter what I do, I will never be good enough for anyone. |
| Online interaction and community | friend, talk, conversation, online, message, post, good, text, reddit, comment, contact, response, advice | I really need someone who can understand and help me with my situation, but I don't know how to ask for help. If anyone wants to message me or comment to my post, I will gladly talk more about it. |
| Family | family, parent, birthday, kid, life, mom, young, age, child, mother, brother, dad | I plan to be homeless next year because I just realized that my family is toxic, and they are damaging my self-esteem. I can't stand anymore. |

| Topic Label | Sample Keywords | Sample Post |
|---|---|---|
| Love | love, heart, pain, dream, moment, deep, touch, happy | I want someone to hug and hold me and make me feel that I am loved. I want to feel safe and real in someone’s arms. |
| Hope for future | life, good, hope, forever, future, change, worth, live, chance, choice | Life is hard but I have always told myself that lots of things can change in a lifetime, hoping for a better tomorrow. But it just slowly keeps getting worse. |
| Job | work, job, back, money, car, city, small, hour, apartment, coworker, plan | It upsets to see myself in early 20s stuck at a dead-end part-time job with no future while everyone else is achieving what they want. No girl would like a loser like me. |
| Hobby | game, good, interest, video, fun, movie, music, great, play interesting, cool, funny | Having hobbies is great but it does not fill the void of loneliness. Life is dull without any friends. |
| College life | friend, school, group, close, college, class, summer, grade, party, university | I live in dorms and just realized how terribly lonely Halloween would be. My family and high school friends will ask me if I am doing anything for Halloween, and I would have to lie again. |
| Human mind | world, human, mind, reality, dark, thought, life, word, sense, deep, moment, truth | My mind during a regular day goes something like this. That girl over there thinks you are ugly, and a group of people over there thinks you are weird. |
| Thoughts | thought, reason, feeling, mind, point, afraid, situation | Does anyone else ever think “what if?” I know past wouldn’t ever be changed, but I keep reflecting the past mistakes and feel worthless. |
| Time period | day, night, week, today, hour, time, back, morning, ago, finally, tonight, late | Tonight is different than the rest of the busy days. I am realizing how lonely and empty I am, which makes me lose appetite and want to fall asleep, but nothing makes me feel tired. |

*Note*. To protect user privacy, Table 5 demonstrates recreated sample posts for each topic.

**Table 6**. Number of posts in common topics between T1 and T2

| Category | Topic | count | | chi-square |
|---|---|---|---|---|
| Mental Health | mental health conditions | T1 | N=632 (5.11%) | 0.53 |
| | | T2 | N=1,761 (5.28%) | |
| | experiencing negative emotions | T1 | N=718 (5.80%) | 29.70* |
| | | T2 | N=2,417 (7.25%) | |
| | venting negative emotions | T1 | N=645 (5.21%) | 15.55* |
| | | T2 | N=2,061 (6.19%) | |
| Social Interaction | relationship | T1 | N=546 (4.41%) | 6.38* |
| | | T2 | N=1,660 (4.98%) | |
| | social evaluation | T1 | N=660 (5.33%) | 0.00 |
| | | T2 | N=1,774 (5.32%) | |
| | social experience | T1 | N=528 (4.27%) | 0.01 |
| | | T2 | N=1,417 (4.25%) | |
| | dating | T1 | N=1,497 (12.10%) | 317.28* |
| | | T2 | N=2,305 (6.92%) | |
| | online interaction and community | T1 | N=726 (5.87%) | 99.54* |
| | | T2 | N=2,903 (8.71%) | |
| Family | family | T1 | N=654 (5.29%) | 11.72* |
| | | T2 | N=2,047 (6.14%) | |
| Emotion | love | T1 | N=724 (5.85%) | 19.43* |
| | | T2 | N=2,335 (7.01%) | |
| | hope for future | T1 | N=576 (4.66%) | 1.42 |
| | | T2 | N=1,642 (4.93%) | |
| Job | job | T1 | N=907 (7.33%) | 45.59* |
| | | T2 | N=1,876 (5.63%) | |

| Category | Topic | count | | chi-square |
|---|---|---|---|---|
| Hobby | hobby | T1 | N=666 (5.38%) | 8.71* |
| | | T2 | N=1,569 (4.71%) | |
| College Life | college life | T1 | N=951 (7.69%) | 0.00 |
| | | T2 | N=2,566 (7.70%) | |
| Human Mind | human mind | T1 | N=687 (5.55%) | 7.91* |
| | | T2 | N=1,632 (4.90%) | |
| | thoughts | T1 | N=411 (3.32%) | 0.33 |
| | | T2 | N=1,144 (3.43%) | |
| Time Period | time period | T1 | N=844 (6.82%) | 0.47 |
| | | T2 | N=2,212 (6.64%) | |

*$p<0.05$

This may be due to the availability of existing subreddits dedicated to mental health conditions (e.g., r/depression)[39] where users can access disease-specific information and support.[32] However, given the significant increases in experiencing and venting negative emotions (e.g., loneliness, sadness, or being depressed), the posts from the mental health category may contain linguistic markers about early signs or symptoms of undiagnosed mental health problems.[70,71]

Under the social interaction category, online interaction and community had a significant increase in DP, reflecting increased use of digital communication during the pandemic.[72,73] Furthermore, given that these posts are related to depression, several terms with high estimated term frequency within the online interaction and community topic (friend, vent, Reddit, advice, etc.) suggest the availability of social support for mental health on online communities such as loneliness subreddits.[32] The DP in a relationship also increased during the pandemic, suggesting the impact of mental health on relationship quality during the pandemic.[74,75]

Moreover, a significant decrease in DP on dating may reflect a decreased engagement in relationship seeking during the pandemic. Given the heightened mental health risks during the pandemic,[76] depression, a health condition characterized by social withdrawal[77] may have affected relationship-seeking behaviors. In fact, we found a shift in dominant topics across two time periods from dating (T1) to online interaction and community (T2). Given the availability of social support on mental health reddit communities,[32] a shift in dominant topic to online interaction and community reflected an increased interest or need in online social support. We found no changes in DPs in social evaluation and social experience despite the disruption in social interaction during the pandemic.[78]

We also found an increased DP around loneliness and depression involving family (family category) during the pandemic. Furthermore, under emotion category, although DPs concerning hope for future showed no difference between T1 and T2, DPs on love had a significant increase during the pandemic. Given the heightened need for mental health support during the pandemic,[79] significant increase may indicate increased expressions or needs in empathetic support for mental health.[80–82]

Moreover, DPs from job category decreased significantly during the pandemic. Although the pandemic has resulted in job insecurity and associated mental health effects (depression, anxiety, etc.),[83] a significant decrease in DPs on job is in contrast with recent studies.[84,85] This may be due to the demographic characteristics of Reddit users where young adults, the primary user group of Reddit,[17] are likely to be in the education system or less likely to be a primary income earner for their household. We also found a decrease in DPs in hobby category. Given that depression is characterized by loss of interest or pleasure in hobbies,[86] our finding highlights the substantial impact of pandemic on mental health.

Interestingly, the college life category showed no changes in DPs. Despite the negative mental health consequences of the COVID-19 pandemic on students,[87–90] DPs did not increase, suggesting school life may not be their main concern. Finally, although the number of DPs in human mind (human mind category) decreased significantly, DPs in thoughts (human mind category) and time period category showed no difference between T1 and T2.

## Discussion

Since the pandemic started, we found an increasing number of posts and authors discussing depression-related topics on loneliness subreddits. We found significant increases in DPs on mental health (i.e., experiencing and venting negative emotions), family, social interaction (i.e., relationship and online interaction and community), and emotion (i.e., love).

We also observed a shift in dominant topics from dating to online interaction and community after the pandemic started, reflecting an increased expression, interest, or need in social support[80–82] given the availability of social support for mental health on online communities[32,91] such as loneliness subreddits. In sum, our study highlights the potential of large-scale naturalistic social media data in reflecting a shift in the mental health experiences of the public in response to an environmental change such as the COVID-19 pandemic.

Compared with specialized mental health discussion boards (e.g., depression subreddits), the loneliness subreddits serves as a more casual and broad discussion board to include discussions on multiple topics beyond mental health conditions. Although no change in DPs on mental health conditions in the loneliness subreddits was found during the pandemic, we found significant increases in experiencing and venting negative emotions, which could be early signs or symptoms of undiagnosed mental health problems.[70,71] Future study should closely examine the linguistic markers for mental health issues from these posts.

Increased DPs from the family category aligns with heightened levels of family conflict during the pandemic.[92–95] Family conflict is closely associated with an increased sense of loneliness[96] and risk for emotional distress, including depressive symptoms.[9] Several studies also reported increased family health concerns during the pandemic, such as concerns around family members experiencing mental health problems and fear or grief of losing a family member.[97] Another significant increase in DPs is situated around relationship and online interaction and community.

Indeed, perceived stress is positively associated with relationship instability during the pandemic.[74] A recent literature also showed an increase in depressive symptoms among individuals with poor relationship quality from 13 to 35 percent during the COVID-19 lockdown.[75] Moreover, a significant increase in online interaction and community may reflect a need for mental health support during the pandemic, given the availability of social support in mental health online communities,[32] whereas difficulties with meeting a demand for mental health treatment were reported among psychologists.[79] In fact, a substantial increase in social media use is associated with feelings of social support.[98]

The results of increasing discussion posts in part reflected an overall spike in social media activity during the pandemic.[99–101] Moreover, a sharp increase in these activities may be related to an increased level of depression among individuals who have been experiencing loneliness-inducing conditions during the pandemic.[15,102–105]

Such results are consistent with higher levels of emotional distress,[69,106] mental health problems emerging from family conflict,[92–95] concerns around family health,[97,107] worsened mental health outcomes related to relationship instability,[108] increased use of online communication channel,[72,73] and increased expressions or needs in mental health support during the pandemic.[80–82] Mental health care providers could be informed accordingly on assessing their clients' mental health and providing timely support (e.g., stress management, having positive relationships/friendships).

**Infoveillance: social media as a window into public mental health**

Lonely people are likely to engage in mental health discussions or seek mental health support online.[24–27,30] From loneliness-related online platforms, we comprehensively understood the users' views and experiences around mental health experiences during the COVID-19 pandemic. Such findings indicate the feasibility of social media posts in delivering the users' real-time issues, concerns, and needs around mental health induced by the significant changes in environmental health (i.e., pandemic). The findings also suggest social media is sensitive and responsive to reflecting interconnected environmental and individual health.

Our study confirms social media has the potential to serve as a window into public mental health, and surveillance through social media enables public mental health monitoring over time and across different crises.[109–112] The current findings and approaches present promising health surveillance measures and an intervening medium.

## Limitations

There are several limitations of this study. First, our classifier that detects depression-related dialogue cannot identify individuals showing depressive symptoms or those at risk of depression. Clinical validation of longitudinal data from individual users by certificated experts (i.e., clinical psychologists or psychiatrists, mental health counselors, rehabilitation counselors) would be needed. Furthermore, although the impact of the pandemic may vary by individual residence (different policies and preventive measures, etc.), our study is limited to the pandemic experiences of U.S. users. In addition, sociodemographic data (gender, employment status, marital status, etc.) were not provided in this study.

This is because self-identification is not required for Reddit user registration, allowing anonymity practices on Reddit (i.e., throwaway accounts).[22,23] Because the anonymity on Reddit allows users to freely share their mental health expressions with minimum concerns around mental health stigma,[32] Reddit is widely utilized to examine mental health discourse.[32,113–115] Although sociodemographic information was not explicitly provided by the users, our results would best reflect the pandemic experience in young adults aged 18–29 in the United States, given the demographic characteristics of Reddit users.[17,21]

In addition, although machine learning-driven thematic analysis revealed the main themes emerging from a large-scale text corpus, future human subject research could complement our findings. Future studies should also examine DPs across different online platforms to gain a broader outlook on pandemic experiences.

## Conclusion

Our study demonstrated that social media analysis could be utilized for public mental health surveillance during a global health crisis. Our future study will advance the predictions to identify at-risk individuals with clinical validations to understand the mental health pandemic better.

## Data Availability Statement

The data underlying this article are publicly available on subreddits of r/lonely,[33] r/loneliness,[34] and r/ForeverAlone,[35] r/depression,[39] and r/NoStupidQuestions.[40]